\documentclass[runningheads]{llncs}
\usepackage[T1]{fontenc}
\usepackage{import}
\usepackage{graphicx,verbatim}
\usepackage{xcolor}
\usepackage{color}
\usepackage{subcaption}
\usepackage{booktabs}
\usepackage{multirow}
\usepackage{pgfplots, tikz}
\usepackage{bm, amsfonts, amsmath, amssymb, dsfont}
\usepackage{hyperref}
\hypersetup{
    colorlinks=true,
    linkcolor=blue,
    filecolor=magenta,      
    urlcolor=cyan,
}

\usepackage{pgfplots}
\pgfplotsset{compat=1.18}
\usepgfplotslibrary{groupplots}
\usepackage{pgfplotstable}

\makeatletter
\newenvironment{customlegend}[1][]{%
    \begingroup
    \pgfplots@init@cleared@structures
    \pgfplotsset{#1}%
}{%
    \pgfplots@createlegend
    \endgroup
}%
\def\addlegendimage{\pgfplots@addlegendimage}
\definecolor{backg}{RGB}{132, 86, 219} 
\definecolor{pg}{RGB}{86, 154, 219} 
\definecolor{rightv}{RGB}{86, 219, 170} 
\definecolor{leftv}{RGB}{116, 219, 86} 
\definecolor{myo}{RGB}{219, 208, 86} 
\definecolor{hipp}{RGB}{219, 94, 86} 

\begin{document}

\author{
Puru Vaish\thanks{Corresponding author.}\inst{1}\orcidID{0000-0002-5180-5293} \and Amin Ranem\inst{1}\orcidID{0000-0003-0783-6903} \and \\
Felix Meister\inst{2} \and
Tobias Heimann\inst{2} \and
Christoph Brune\inst{1}\orcidID{0000-0003-0145-5069} \and 
Jelmer M.~Wolterink\inst{1}\orcidID{0000-0001-5505-475X}
}
\authorrunning{Vaish et al.}
\institute{
Department of Applied Mathematics, Technical Medical Centre, \\ University of Twente, The Netherlands\\
\email{\{p.vaish, a.ranem, c.brune, j.m.wolterink\}@utwente.nl}
\and
Digital Technology and Innovation, Siemens Healthineers, Erlangen, Germany\\
\email{\{felix.meister, tobias.heimann\}@siemens-healthineers.com}
}


\title{SegReg: Latent Space Regularization for Improved Medical Image Segmentation}
\titlerunning{SegReg}

\maketitle              
\begin{abstract}
Medical image segmentation models are typically optimised with voxel-wise losses that constrain predictions only in the output space. This leaves latent feature representations largely unconstrained, potentially limiting generalisation. We propose {SegReg}, a latent-space regularisation framework that operates on feature maps of U-Net models to encourage structured embeddings while remaining fully compatible with standard segmentation losses. Integrated with the nnU-Net framework, we evaluate SegReg on prostate, cardiac, and hippocampus segmentation and demonstrate consistent improvements in domain generalisation. Furthermore, we show that explicit latent regularisation improves continual learning by reducing task drift and enhancing forward transfer across sequential tasks without adding memory or any extra parameters. These results highlight latent-space regularisation as a practical approach for building more generalisable and continual-learning-ready models.
\end{abstract}

\section{Introduction}\label{sec:intro}

Deep learning is the dominant paradigm for medical image segmentation, where encoder–decoder architectures such as U-Net~\cite{ronneberger_u-net_2015} learn dense voxel-wise predictions. State-of-the-art methods~\cite{isensee_nnu-net_2021} optimise voxel-level objectives such as Dice or cross-entropy loss, constraining predictions in the output space. While effective, these objectives provide little explicit control over the structure of intermediate feature representations. Consequently, latent spaces emerge implicitly from optimisation rather than being deliberately shaped.

\begin{figure}[tb]
    \centering
    \def\svgwidth{0.97\linewidth}
    \import{figures/schema}{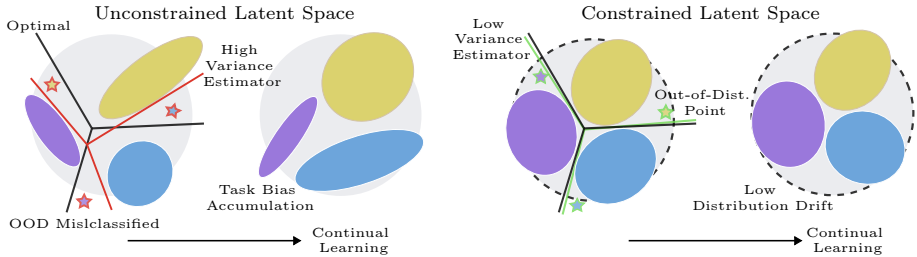}
\caption{\textit{SegReg} introduces explicit latent-space regularisation that aligns embeddings with a fixed reference distribution, thereby reducing classifier variability. By stabilising the latent space, SegReg improves domain generalisation and furthermore for continual learning, anchoring representations to a fixed reference distribution mitigates bias accumulation across sequential training stages.}
    \label{fig:schema}
\end{figure}

Regularisation is central to improving generalisation in deep networks. In medical segmentation, it is typically implicit: architectural inductive biases (e.g., U-Net, residual learning~\cite{he_deep_2016}), automated design choices in nnU-Net~\cite{isensee_nnu-net_2021}, data augmentation~\cite{shorten_survey_2019,vaish_data-agnostic_2025}, adversarial training~\cite{dou_unsupervised_2018}, and self-supervised learning~\cite{chen_simple_2020} indirectly shape learned embeddings.

An alternative approach \emph{explicitly} regularises the latent spaces. Metric learning enforces structured embeddings through contrastive or triplet objectives~\cite{hadsell_dimensionality_2006,chen_simple_2020}, while probabilistic models such as variational autoencoders impose distributional priors~\cite{kingma_auto-encoding_2014}. In medical imaging, latent constraints appear as anatomical priors~\cite{oktay_anatomically_2018}, probabilistic latent modelling~\cite{kohl_probabilistic_2018}, or contrastive feature regularisation~\cite{chaitanya_contrastive_2020}. However, these approaches typically enforce relative relationships (e.g., class similarity or domain alignment) or task-specific priors, rather than constraining the global structure of the latent distribution itself.

This motivates a fundamental question: \emph{is there a reference distribution in latent space that segmentation models should optimise for?} If such a reference exists, enforcing it could provide structured and stable representations, reduce overfitting, and improve domain generalisation. Several works hint at this possibility. For example, adversarial feature alignment~\cite{kamnitsas_unsupervised_2017}, domain-invariant learning~\cite{dou_unsupervised_2018}, and entropy minimisation~\cite{vu_advent_2019} align latent spaces across domains. However, they primarily address domain adaptation rather than shaping representation geometry during standard supervised training.

We argue that latent-space regularisation should not be defined solely by static training objectives but should also account for the lifecycle of an image segmentation model. Medical segmentation systems are increasingly deployed in evolving clinical settings, where natural distribution shifts occur over time, e.g., due to changes in imaging protocols or technologies. Under such conditions, learned representations must remain stable while adapting to new distributions - a setting naturally described by continual learning. Continual learning studies how neural networks can acquire new tasks without catastrophic forgetting~\cite{mccloskey_catastrophic_1989,kirkpatrick_overcoming_2017}. In medical imaging, recent works demonstrate the challenge of retaining segmentation knowledge across sequential datasets and anatomies~\cite{gonzalez_lifelong_2023}. Yet most approaches focus on parameter-space constraints~\cite{kirkpatrick_overcoming_2017} or replay mechanisms~\cite{rolnickExperienceReplayContinual2019}, leaving the dynamics of latent representations underexplored.

In this work, we propose {SegReg}, a latent-space regularisation framework that explicitly constrains penultimate-layer feature maps of U-Net models. Integrated with (Lifelong) nnU-Net framework~\cite{isensee_nnu-net_2021,gonzalez_lifelong_2023}, SegReg enforces a reference latent distribution without modifying the segmentation architecture or introducing additional parameters with minimal training overhead, enabling evaluation for both static and continual learning settings.

Our contributions are threefold. First, we introduce an explicit latent-space regularisation framework for medical image segmentation and formalise the problem as learning a reference latent distribution. Second, we demonstrate consistent improvements in segmentation performance and domain generalisation across multiple benchmarks under standardized (Lifelong) nnU-Net training framework. Third, we show that the same latent constraints stabilise representations in continual learning, reducing task drift and improving forward transfer without replay buffers or additional model parameters.
\section{Methodology}\label{sec:method}

\subsection{Latent Distribution Drift}
To motivate our methodology, we must first model the evolution of latent representations across sequential training stages as a way to understand and mitigate feature instability. Let $\mathcal{S}=\{1,\dots,T\}$ denote a sequence of tasks, where each stage $t$ introduces a dataset $(\mathbf{x},y) \sim p_t(\mathbf{x},y)$, with $\mathbf{x}$ representing input images and $y$ their segmentation labels. Rather than analysing learning solely through parameter updates, we characterise how the encoder $f_\theta:\mathcal{X}\rightarrow\mathcal{Z}$ maps inputs to latent representations $\mathbf{z}=f_\theta(\mathbf{x}) \in \mathbb{R}^d$ across tasks.

For each task $t$, the input marginal and its induced latent distribution are:
\begin{align}
p_t(\mathbf{x})
= \int p_t(\mathbf{x}, y)\,dy, &&
q_t(\mathbf{z})
= \int_{\mathcal{X}}
p_t(\mathbf{x})\,
\delta\!\left(\mathbf{z}-f_\theta(\mathbf{x})\right)
\,d\mathbf{x},
\end{align}
where $\delta(\cdot)$ denotes the Dirac delta distribution.

Instability or drift in these latent representations can degrade performance on previously learned tasks. One way to formalise this is through the Kullback-Leibler (KL) divergence between consecutive latent distributions~\cite{cover_differential_2005}:
\begin{equation}
\mathcal{R}_T =
\sum_{t=2}^{T}
D_{\mathrm{KL}}(q_t \,\|\, q_{t-1})
=
\sum_{t=2}^{T}
\int_{\mathcal{Z}}
q_t(\mathbf{z})
\log \frac{q_t(\mathbf{z})}{q_{t-1}(\mathbf{z})}
\, d\mathbf{z}.
\end{equation}

In practice, however, directly minimising $\mathcal{R}_T$ is infeasible: evaluating the KL divergence requires access to high-dimensional latent distributions of past tasks, which is intractable without storing previous data or fully converged models. This motivates the introduction of a fixed reference latent distribution toward which each task’s embeddings should be regularised. By aligning latent representations to a fixed target distribution, we reduce representational drift. This raises the question, \textit{which reference distribution should be chosen?}

\paragraph{Reference Latent Distributions under Moment Constraints.}
Assume latent representations are constrained to have zero mean and fixed covariance $\Sigma \succ 0$. Let $\mathcal{P}_\Sigma$ denote the set of all distributions with these moments. We seek a reference
distribution $r \in \mathcal{P}_\Sigma$ that yields stable behaviour under distributional shifts.

\begin{theorem}[Extremal Property of the Gaussian Distribution]
\label{thm:gaussian_opt}
Let $r \in \mathcal{P}_{\Sigma}$ and $g \sim \mathcal{N}(\mathbf{0},\Sigma)$. Then
$h(r)\leq h(g)$, where $h(\cdot)$ denotes differential entropy.
\end{theorem}

The proof follows directly from Cover, et. al.~\cite{cover_differential_2005}. Importantly, the result does not merely state that the Gaussian has maximum entropy. Rather, it implies that under fixed second-order statistics, the Gaussian corresponds to the \emph{least informative} distribution, i.e., the one that introduces minimal additional structure beyond the imposed constraints. Balestriero et. al.~\cite{balestriero_lejepa_2025} also proved that isotropic Gaussian distributed latents result in minimum bias estimator. Regularising latent representations toward such a distribution avoids imposing task-specific biases, promotes lower variance estimator, and the worst-case divergence between stages is reduced under perturbations or task shifts making it the optimal choice.

\subsection{Regularisation Components}
\label{sec:sigreg_inv}
We adopt SIGReg~\cite{balestriero_lejepa_2025}, which formulates a statistical test on the distribution of random one-dimensional projections of latent embeddings toward an isotropic Gaussian. In the segmentation setting, we replace the original multiview invariance objective with a supervised, class-wise aggregation scheme. During training, each voxel embedding is regularised towards the prototype of its corresponding ground-truth class. The SIGReg term employs the Epps--Pulley test, $\operatorname{T}$, to encourage well-conditioned feature statistics, while the invariance term promotes intra-class compactness around global class means.

Formally, consider a task with $C$ classes and $N$ voxels in a batch. Let $\boldsymbol{z}_i \in \mathbb{R}^d$ denote the latent embedding of voxel $i$. Define $\mathcal{A} = \{\mathbf{a} \in \mathbb{R}^{d} : \|a\|_2 = 1 \}$ to be the set of random projection vectors. For each class $c \in \{0, \ldots, C\}$, define the index set of voxels belonging to that class as $\mathcal{I}_c$. The class prototype is defined as $ \boldsymbol{\mu}_c = \frac{1}{|\mathcal{I}_c|} \sum_{i \in \mathcal{I}_c} \boldsymbol{z}_i.$ Then the SIGReg and invariance losses are defined as:
\begin{align}
\mathcal{L}_{\text{SIGReg}}
= \frac{1}{|\mathbb{A}|}
\sum_{\boldsymbol{a} \in \mathbb{A}}
\operatorname{T}\!\left(\left\{\boldsymbol{a}^{\top} \boldsymbol{z}_i\right\}_{i=1}^N\right), &&
\mathcal{L}_{\text{Inv}}
= \frac{1}{C}
\sum_{c=1}^{C}
\frac{1}{|\mathcal{I}_c|}
\sum_{i \in \mathcal{I}_c}
\left\| \boldsymbol{z}_i - \boldsymbol{\mu}_c \right\|_2^2 .
\end{align}

\subsection{Proposed Method: SegReg}
Building on the latent regularisation components described above, we propose \textit{SegReg}. SegReg does not replace standard segmentation objectives; instead, it combines them with latent-space constraints that promote statistically well-conditioned and class-consistent embeddings.

For any base segmentation objective, $\mathcal{L}_{\text{Seg}}$, the overall training objective is:
\begin{equation}
\mathcal{L}_{\text{SegReg}}
=
\mathcal{L}_{\text{Seg}}
+ \lambda\, \mathcal{L}_{\text{SIGReg}}
+ (1-\lambda)\, \mathcal{L}_{\text{Inv}},
\end{equation}
where $\mathcal{L}_{\text{Seg}}$ denotes the base segmentation loss. The weighting parameter $\lambda \in [0,1]$ controls the trade-off between distributional regularisation and invariance constraints. By jointly imposing distributional conditioning and class-wise invariance, SegReg stabilises the latent manifold across training iterations and across training stages in continual learning reducing representational drift while preserving segmentation performance.

\section{Experiments}\label{sec:exp}

\paragraph{Datasets.}
We evaluate SegReg on multiple medical image segmentation datasets covering diverse anatomies and imaging distributions. For prostate segmentation, we use a multi-site dataset~\cite{liuMSnetMultisiteNetwork2020}: UCL (13 cases), I2CVB (19 cases), ISBI (30 cases), and Decathlon Prostate (DecP, 32 cases)~\cite{simpson_large_2019}. Cardiac segmentation uses scanner-specific subsets of MnMs~\cite{campelloMulticentreMultivendorMultidisease2021}: Siemens (hA, 75 cases) and Philips (hB, 75 cases), as well as the CMR dataset (160 cases) for static training. Hippocampus segmentation is evaluated on HarP (270 cases)~\cite{boccardi_training_2015}, Dryad (50 cases)~\cite{kulaga-yoskovitz_multi-contrast_2015}, and Hippocampus Decathlon (DecH, 260 cases)~\cite{simpson_large_2019}. These datasets differ in anatomy, resolution, and acquisition protocols, enabling systematic evaluation of domain generalisation and continual learning.

\paragraph{Training.} All experiments use the nnU-Net~\cite{isensee_nnu-net_2021} framework to ensure reproducibility and a strong baseline. Unless otherwise stated, all methods are trained and evaluated using identical dataset splits, we adopt the default preprocessing, augmentation, optimisation, and hyperparameters automatically configured by nnU-Net for each dataset. For continual learning, we follow the Lifelong nnU-Net protocol~\cite{gonzalez_lifelong_2023}, including the predefined task sequences and evaluation procedure. SegReg uses a fixed $\lambda=0.05$ chosen based on a small ablation. SegReg does not introduce additional trainable parameters or buffers during training.

\paragraph{Metrics.} For domain generalisation, we report Dice similarity coefficient (DSC) when training on one dataset and testing on others using \texttt{MONAI} implementation~\cite{consortiumMONAIMedicalOpen2026}. For continual learning, we follow~\cite{delangeContinualLearningSurvey2022} and report: (i) mean task performance at the end of training, measured as mean Dice ($\bar{\mathrm{DSC}}$) across all tasks; (ii) mean forward transfer ($\bar{\mathrm{FWT}}$), defined as difference in performance on a task before it is learned and training on the task independently; (iii) mean backward transfer ($\bar{\mathrm{BWT}}$), quantifying performance changes on previously learned tasks.

\subsection{Results}
\begin{table}[t]
	\centering
	\caption{
        DSC scores comparing nnU-Net trained with and without SegReg. Each row is a training dataset, columns show test results per method along with standard deviation. The last column shows the average domain generalisation improvement per training dataset. \textbf{Bold} face numbers, under the paired t-test with $p < 0.05$, indicate results rejecting the null hypothesis of no improvement.
    }
	\label{tab:cross_domain}
    {
        \fontsize{8pt}{10pt}\selectfont
        \begin{tabular}{
@{}
l
*{24}{c}
r
@{}
}
\toprule
\textbf{Train} & \multicolumn{12}{c}{\textbf{SegReg}}                                                                                                                                                      & \multicolumn{12}{c}{\textbf{Baseline}}                                                                                                                                                   & $\Bar{\Delta}$ \\ \cmidrule(r){1-1}\cmidrule(lr){2-13}\cmidrule(lr){14-25}\cmidrule(l){26-26}
\textbf{Prostate} & \multicolumn{3}{c}{UCL}                      & \multicolumn{3}{c}{I2CVB}                    & \multicolumn{3}{c}{ISBI}                     & \multicolumn{3}{c}{DecP}                     & \multicolumn{3}{c}{UCL}                      & \multicolumn{3}{c}{I2CVB}                    & \multicolumn{3}{c}{ISBI}                     & \multicolumn{3}{c}{DecP}                    & 8.4            \\ \cmidrule(r){1-1}\cmidrule(lr){2-4}\cmidrule(lr){5-7}\cmidrule(lr){8-10}\cmidrule(lr){11-13}\cmidrule(lr){14-16}\cmidrule(lr){17-19}\cmidrule(lr){20-22}\cmidrule(lr){23-25}\cmidrule(l){26-26}

DecP           & \multicolumn{3}{c}{\textbf{76.7}$\pm${ 2}}  & \multicolumn{3}{c}{\textbf{48.5}$\pm${30}} & \multicolumn{3}{c}{\textbf{87.0}$\pm${ 5}}  & \multicolumn{3}{c}{\textbf{90.7}$\pm${ 3}}  & \multicolumn{3}{c}{67.3$\pm${ 3}}  & \multicolumn{3}{c}{10.2$\pm${14}} & \multicolumn{3}{c}{79.3$\pm${ 6}}  & \multicolumn{3}{c}{89.6$\pm${ 4}} & 14             \\
ISBI           & \multicolumn{3}{c}{\textbf{78.9}$\pm${ 2}}  & \multicolumn{3}{c}{\textbf{64.6}$\pm${23}} & \multicolumn{3}{c}{92.8$\pm${ 1}}  & \multicolumn{3}{c}{\textbf{90.3}$\pm${ 3}}  & \multicolumn{3}{c}{78.0$\pm${ 3}}  & \multicolumn{3}{c}{51.6$\pm${37}} & \multicolumn{3}{c}{92.5$\pm${ 1}}  & \multicolumn{3}{c}{86.4$\pm${ 3}} & 4.5            \\
I2CVB          & \multicolumn{3}{c}{28.8$\pm${29}} & \multicolumn{3}{c}{\textbf{84.7}$\pm${ 1}}  & \multicolumn{3}{c}{\textbf{23.6}$\pm${11}} & \multicolumn{3}{c}{~2.9$\pm${ 5}}   & \multicolumn{3}{c}{25.7$\pm${26}} & \multicolumn{3}{c}{83.9$\pm${ 3}}  & \multicolumn{3}{c}{12.7$\pm${14}} & \multicolumn{3}{c}{~0.8$\pm${ 5}} & 4.2            \\
UCL            & \multicolumn{3}{c}{\textbf{82.0}$\pm${ 6}}  & \multicolumn{3}{c}{\textbf{39.9}$\pm${16}} & \multicolumn{3}{c}{\textbf{69.6}$\pm${22}} & \multicolumn{3}{c}{\textbf{58.9}$\pm${13}} & \multicolumn{3}{c}{76.6$\pm${ 9}}  & \multicolumn{3}{c}{17.3$\pm${30}} & \multicolumn{3}{c}{63.3$\pm${27}} & \multicolumn{3}{c}{50.1$\pm${17}} & 11             \\ \midrule
\textbf{Cardiac} & \multicolumn{4}{>{\centering\arraybackslash}p{0.13\linewidth}}{hB}                                        & \multicolumn{4}{>{\centering\arraybackslash}p{0.13\linewidth}}{hA}                                      & \multicolumn{4}{>{\centering\arraybackslash}p{0.13\linewidth}}{CMR}                                     & \multicolumn{4}{>{\centering\arraybackslash}p{0.13\linewidth}}{hB}                                        & \multicolumn{4}{>{\centering\arraybackslash}p{0.13\linewidth}}{hA}                                      & \multicolumn{4}{>{\centering\arraybackslash}p{0.13\linewidth}}{CMR}                                    & 1.7            \\ \cmidrule(r){1-1}\cmidrule(lr){2-5}\cmidrule(lr){6-9}\cmidrule(lr){10-13}\cmidrule(lr){14-17}\cmidrule(lr){18-21}\cmidrule(lr){22-25}\cmidrule(l){26-26}

CMR            & \multicolumn{4}{>{\centering\arraybackslash}p{0.13\linewidth}}{88.1$\pm${ 5}}                   & \multicolumn{4}{>{\centering\arraybackslash}p{0.13\linewidth}}{73.4$\pm${14}}                & \multicolumn{4}{>{\centering\arraybackslash}p{0.13\linewidth}}{73.4$\pm${32}}                & \multicolumn{4}{>{\centering\arraybackslash}p{0.13\linewidth}}{88.2$\pm${ 5}}                   & \multicolumn{4}{>{\centering\arraybackslash}p{0.13\linewidth}}{73.3$\pm${17}}                & \multicolumn{4}{>{\centering\arraybackslash}p{0.13\linewidth}}{72.3$\pm${32}}               & 0.4            \\
hA             & \multicolumn{4}{>{\centering\arraybackslash}p{0.13\linewidth}}{88.6$\pm${ 4}}                   & \multicolumn{4}{>{\centering\arraybackslash}p{0.13\linewidth}}{89.1$\pm${ 5}}                 & \multicolumn{4}{>{\centering\arraybackslash}p{0.13\linewidth}}{\textbf{66.2}$\pm${30}}                & \multicolumn{4}{>{\centering\arraybackslash}p{0.13\linewidth}}{88.0$\pm${ 4}}                   & \multicolumn{4}{>{\centering\arraybackslash}p{0.13\linewidth}}{88.4$\pm${ 6}}                 & \multicolumn{4}{>{\centering\arraybackslash}p{0.13\linewidth}}{63.7$\pm${30}}               & 1.3            \\
hB             & \multicolumn{4}{>{\centering\arraybackslash}p{0.13\linewidth}}{92.5$\pm${ 3}}                   & \multicolumn{4}{>{\centering\arraybackslash}p{0.13\linewidth}}{\textbf{53.9}$\pm${30}}                & \multicolumn{4}{>{\centering\arraybackslash}p{0.13\linewidth}}{70.1$\pm${31}}                & \multicolumn{4}{>{\centering\arraybackslash}p{0.13\linewidth}}{92.4$\pm${ 3}}                   & \multicolumn{4}{>{\centering\arraybackslash}p{0.13\linewidth}}{44.3$\pm${30}}                & \multicolumn{4}{>{\centering\arraybackslash}p{0.13\linewidth}}{69.6$\pm${31}}               & 3.6            \\ \midrule
\textbf{Hipp.} & \multicolumn{4}{>{\centering\arraybackslash}p{0.13\linewidth}}{HarP}                                      & \multicolumn{4}{>{\centering\arraybackslash}p{0.13\linewidth}}{Dryad}                                   & \multicolumn{4}{>{\centering\arraybackslash}p{0.13\linewidth}}{DecH}                                    & \multicolumn{4}{>{\centering\arraybackslash}p{0.13\linewidth}}{HarP}                                      & \multicolumn{4}{>{\centering\arraybackslash}p{0.13\linewidth}}{Dryad}                                   & \multicolumn{4}{>{\centering\arraybackslash}p{0.13\linewidth}}{DecH}                                   & 4.3            \\ \cmidrule(r){1-1}\cmidrule(lr){2-5}\cmidrule(lr){6-9}\cmidrule(lr){10-13}\cmidrule(lr){14-17}\cmidrule(lr){18-21}\cmidrule(lr){22-25}\cmidrule(l){26-26}
DecH           & \multicolumn{4}{>{\centering\arraybackslash}p{0.13\linewidth}}{\textbf{57.8}$\pm${16}}                  & \multicolumn{4}{>{\centering\arraybackslash}p{0.13\linewidth}}{\textbf{79.4}$\pm${ 4}}                 & \multicolumn{4}{>{\centering\arraybackslash}p{0.13\linewidth}}{90.6$\pm${ 3}}                 & \multicolumn{4}{>{\centering\arraybackslash}p{0.13\linewidth}}{55.6$\pm${18}}                  & \multicolumn{4}{>{\centering\arraybackslash}p{0.13\linewidth}}{56.6$\pm${ 5}}                 & \multicolumn{4}{>{\centering\arraybackslash}p{0.13\linewidth}}{90.6$\pm${ 3}}                & 8.3            \\
Dryad          & \multicolumn{4}{>{\centering\arraybackslash}p{0.13\linewidth}}{34.6$\pm${28}}                  & \multicolumn{4}{>{\centering\arraybackslash}p{0.13\linewidth}}{92.6$\pm${ 1}}                 & \multicolumn{4}{>{\centering\arraybackslash}p{0.13\linewidth}}{\textbf{86.0}$\pm${ 3}}                 & \multicolumn{4}{>{\centering\arraybackslash}p{0.13\linewidth}}{33.2$\pm${31}}                  & \multicolumn{4}{>{\centering\arraybackslash}p{0.13\linewidth}}{92.7$\pm${ 1}}                 & \multicolumn{4}{>{\centering\arraybackslash}p{0.13\linewidth}}{84.7$\pm${ 4}}                & 0.9            \\
HarP           & \multicolumn{4}{>{\centering\arraybackslash}p{0.13\linewidth}}{88.8$\pm${ 9}}                   & \multicolumn{4}{>{\centering\arraybackslash}p{0.13\linewidth}}{\textbf{88.4}$\pm${ 1}}                 & \multicolumn{4}{>{\centering\arraybackslash}p{0.13\linewidth}}{\textbf{42.4}$\pm${15}}                & \multicolumn{4}{>{\centering\arraybackslash}p{0.13\linewidth}}{88.8$\pm${ 9}}                   & \multicolumn{4}{>{\centering\arraybackslash}p{0.13\linewidth}}{87.8$\pm${ 1}}                 & \multicolumn{4}{>{\centering\arraybackslash}p{0.13\linewidth}}{32.1$\pm${17}}               & 3.6            \\ \bottomrule
\end{tabular}

    }
\end{table}

\paragraph{Domain Generalization.}
Tab.~\ref{tab:cross_domain} reports domain segmentation performance across prostate, cardiac, and hippocampus MRI. SegReg consistently improves generalisation over the nnU-Net baseline. For prostate MRI, we observe large gains (mean +8.4 DSC), particularly under strong domain shifts (e.g., DecP$\rightarrow$I2CVB: 10.2 to 48.5). Hippocampus datasets show similarly notable improvements (mean +4.3 DSC), including challenging transfers such as DecH$\rightarrow$Dryad (56.6 to 79.4). Cardiac MRI exhibits smaller but consistent gains (mean +1.7 DSC), with improvements concentrated in hB$\rightarrow$hA (44.3 to 53.9). Across anatomies, improvements are most pronounced when source and target domains differ substantially. 

Overall, the consistent domain-generalisation gains across anatomies suggest that structuring the latent space contributes to improved feature transfer under distribution shift.  While not explicitly enforcing invariance, these results indicate that latent-space regularisation can enhance robustness to unseen domains and promote more stable generalisation across heterogeneous datasets.

\paragraph{Continual Learning.} We evaluate SegReg against standard continual learning methods: sequential fine-tuning without constraints (Seq.), Elastic Weight Consolidation (EWC) which reduces parameter changes between stages, and Random Walk (RWalk). We treat rehearsal with stored samples (Reh.) as an upper baseline as our method falls under the category of regularisation based methods.

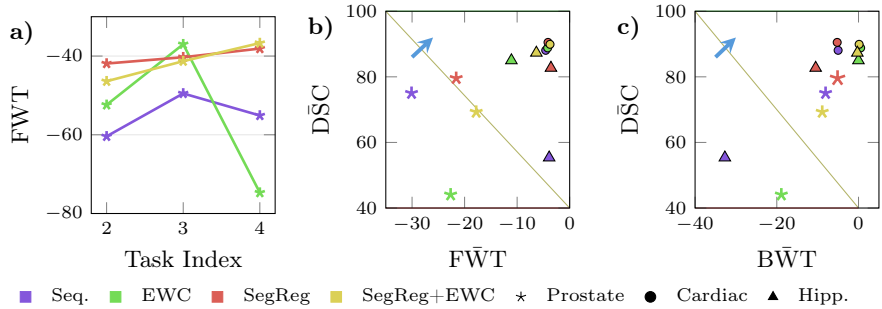
\begin{figure}[tb]
    \centering
    \begin{subfigure}{0.33\textwidth}
\centering
\begin{tikzpicture}
\node[anchor=south west, font=\bfseries] at (-1.2,2.125) {a)};
\begin{axis}[
    width=\linewidth,
    height=4.2cm,
    xlabel={Task Index},
    ylabel={FWT},
    xtick={1,2,3},
    xticklabels={2,3,4},
    ymin=-80, ymax=-30,
    ymajorgrids=true,
    grid style={gray!20},
    tick label style={font=\scriptsize},
    label style={font=\small},
    every axis plot/.append style={line width=1pt},
]
\addplot[color=backg, mark=star] coordinates {(1,-60.4) (2,-49.5) (3,-55.1)};
\addplot[color=leftv, mark=star] coordinates {(1,-52.4) (2,-37.0) (3,-74.7)};
\addplot[color=hipp, mark=star] coordinates {(1,-41.9) (2,-40.3) (3,-38.1)};
\addplot[color=myo, mark=star] coordinates {(1,-46.4) (2,-41.3) (3,-36.7)};
\end{axis}
\end{tikzpicture}
\end{subfigure}%
\begin{subfigure}{0.33\linewidth}
\centering
\begin{tikzpicture}
\node[anchor=south west, font=\bfseries] at (-1.15,2.125) {b)};
\begin{axis}[
    width=\linewidth,
    height=4.2cm,
    xlabel={$\Bar{\text{FWT}}$},
    ylabel={$\Bar{\text{DSC}}$},
    xmin=-35, xmax=0,
    ymin=40, ymax=100,
    enlargelimits=false,
    tick label style={font=\scriptsize},
    label style={font=\small},
    set layers
]

\addplot[
    on layer=axis background,
    mesh,
    mesh/rows=2,
    shader=interp,
    draw=none,
    opacity=0.6,
    colormap={redgreen}{
        color(0cm)=(red);
        color(1cm)=(green)
    }
] coordinates {
    (-35,40)   [0]
    (0,40)     [0.5]
    (-35,100)  [0.5]
    (0,100)    [1]
};
\draw[-stealth, pg, line width=1.5pt] (-30,86) -- (-26,92) 
    node[pos=0.5, above right, font=\large] {};

\addplot[only marks, mark=star, mark size=2.5pt, color=backg, line width=1pt]
coordinates {(-30.10,75.03)};
\addplot[only marks, mark=star, mark size=2.5pt, color=leftv, line width=1pt]
coordinates {(-22.66,44.07)};
\addplot[only marks, mark=star, mark size=2.5pt, color=hipp, line width=1pt]
coordinates {(-21.60,79.58)};
\addplot[only marks, mark=star, mark size=2.5pt, color=myo, line width=1pt]
coordinates {(-17.79,69.28)};

\addplot[only marks, mark=*, mark size=1.5pt, color=backg, draw=black]
coordinates {(-4.56,88.08)};
\addplot[only marks, mark=*, mark size=1.5pt, color=leftv, draw=black]
coordinates {(-4.19,88.82)};
\addplot[only marks, mark=*, mark size=1.5pt, color=hipp, draw=black]
coordinates {(-4.12,90.50)};
\addplot[only marks, mark=*, mark size=1.5pt, color=myo, draw=black]
coordinates {(-3.74,89.91)};

\addplot[only marks, mark=triangle*, mark size=2.5pt, color=backg, draw=black]
coordinates {(-3.90,55.37)};
\addplot[only marks, mark=triangle*, mark size=2.5pt, color=leftv, draw=black]
coordinates {(-11.09,85.00)};
\addplot[only marks, mark=triangle*, mark size=2.5pt, color=hipp, draw=black]
coordinates {(-3.56,82.71)};
\addplot[only marks, mark=triangle*, mark size=2.5pt, color=myo, draw=black]
coordinates {(-6.34,87.32)};
\end{axis}
\end{tikzpicture}
\end{subfigure}%
\begin{subfigure}{0.33\linewidth}
\centering
\begin{tikzpicture}
\node[anchor=south west, font=\bfseries] at (-1.15,2.125) {c)};
\begin{axis}[
    width=\linewidth,
    height=4.2cm,
    xlabel={$\Bar{\text{BWT}}$},
    ylabel={$\Bar{\text{DSC}}$},
    xmin=-40, xmax=5,
    ymin=40, ymax=100,
    enlargelimits=false,
    tick label style={font=\scriptsize},
    label style={font=\small},
    set layers
]

\addplot[
    on layer=axis background,
    mesh,
    mesh/rows=2,
    shader=interp,
    draw=none,
    opacity=0.6,
    colormap={redgreen}{
        color(0cm)=(red);
        color(1cm)=(green)
    }
] coordinates {
    (-40,40)   [0]
    (0,40)     [0.5]
    (-40,100)  [0.5]
    (0,100)    [1]
};
\draw[-stealth, pg, line width=1.5pt] (-35,86) -- (-30,92) 
    node[pos=0.5, above right, font=\scriptsize] {};

\addplot[only marks, mark=star, mark size=2.5pt, color=backg, line width=1pt]
coordinates {(-8.06,75.03)};
\addplot[only marks, mark=star, mark size=2.5pt, color=leftv, line width=1pt]
coordinates {(-18.89,44.07)};
\addplot[only marks, mark=star, mark size=3pt, color=hipp, line width=1pt]
coordinates {(-5.13,79.58)};
\addplot[only marks, mark=star, mark size=2.5pt, color=myo, line width=1pt]
coordinates {(-8.88,69.28)};

\addplot[only marks, mark=*, mark size=1.5pt, color=backg, draw=black]
coordinates {(-4.99,88.08)};
\addplot[only marks, mark=*, mark size=1.5pt, color=leftv, draw=black]
coordinates {(0.52,88.82)};
\addplot[only marks, mark=*, mark size=1.5pt, color=hipp, draw=black]
coordinates {(-5.22,90.50)};
\addplot[only marks, mark=*, mark size=1.5pt, color=myo, draw=black]
coordinates {(0.15,89.91)};

\addplot[only marks, mark=triangle*, mark size=2.5pt, color=backg, draw=black]
coordinates {(-32.71,55.37)};
\addplot[only marks, mark=triangle*, mark size=2.5pt, color=leftv, draw=black]
coordinates {(-0.02,85.00)};
\addplot[only marks, mark=triangle*, mark size=2.5pt, color=hipp, draw=black]
coordinates {(-10.49,82.71)};
\addplot[only marks, mark=triangle*, mark size=2.5pt, color=myo, draw=black]
coordinates {(-0.28,87.32)};
\end{axis}
\end{tikzpicture}
\end{subfigure}

\begin{tikzpicture}
\begin{customlegend}[
    legend columns=-1,
    legend style={draw=none,column sep=1.5ex, font=\scriptsize},
    legend entries={Seq., EWC, SegReg, SegReg+EWC, Prostate, Cardiac, Hipp.}
]
\addlegendimage{only marks,mark=square*,color=backg}
\addlegendimage{only marks,mark=square*,color=leftv}
\addlegendimage{only marks,mark=square*,color=hipp}
\addlegendimage{only marks,mark=square*,color=myo}
\addlegendimage{only marks,mark=star}
\addlegendimage{only marks,mark=*}
\addlegendimage{only marks,mark=triangle*}
\end{customlegend}
\end{tikzpicture}
    \caption{a) Forward transfer as a function of task index for prostate datasets. b–c) Shows the relationship between $\Bar{\text{DSC}}$ and $\Bar{\text{FWT}}$/$\Bar{\text{BWT}}$ respectively where the improvement towards top-right shows improved transfer and performance.}
    \label{fig:cl_metrics}
\end{figure}

Fig.~\ref{fig:cl_metrics} shows knowledge transfer dynamics in the continual learning setting. Fig.~\ref{fig:cl_metrics}(a) shows forward transfer (FWT) across tasks on the prostate sequence. SegReg shows consistently improving forward transfer as tasks progress, whereas Seq. and EWC exhibit unstable trends and a marked drop on later tasks. Figures~\ref{fig:cl_metrics}(b–c) relate final mean Dice to transfer metrics. SegReg achieves a superior $\Bar{\text{FWT}}$–$\Bar{\text{DSC}}$ trade-off, indicating more effective reuse of previously acquired knowledge. Backward transfer remains competitive and is highest on the prostate setting while its combination with EWC consistently improves forward transfer, showing that improved FWT does not come at the expense of retention.

\begin{table}[tb]
    \centering
    \caption{Continual learning performance metrics ($\Bar{\text{DSC}}$, $\Bar{\text{FWT}}$, $\Bar{\text{BWT}}$) for Prostate, Cardiac, and Hippocampus datasets along with the std. over tasks. The best results are highlighted in \textbf{bold}, and the second-best results are \underline{underlined}.}
    \label{tab:cl_metrics}
    {
        \fontsize{8pt}{10pt}\selectfont  
        \begin{tabular}{l@{\hspace{2.5mm}}ccc@{\hspace{2.5mm}}ccc@{\hspace{2.5mm}}ccc}
\toprule
\multirow{2}{*}{Method} & \multicolumn{3}{c}{Prostate} & \multicolumn{3}{c}{Cardiac} & \multicolumn{3}{c}{Hippocampus} \\ \cmidrule(lr){2-4} \cmidrule(lr){5-7} \cmidrule(l){8-10}
 & $\Bar{\text{DSC}}$ & $\Bar{\text{BWT}}$ & $\Bar{\text{FWT}}$ & $\Bar{\text{DSC}}$ & $\Bar{\text{BWT}}$ & $\Bar{\text{FWT}}$ & $\Bar{\text{DSC}}$ & $\Bar{\text{BWT}}$ & $\Bar{\text{FWT}}$ \\ \cmidrule(r){1-1} \cmidrule(lr){2-2} \cmidrule(lr){3-3} \cmidrule(lr){4-4} \cmidrule(lr){5-5} \cmidrule(lr){6-6} \cmidrule(lr){7-7} \cmidrule(lr){8-8} \cmidrule(lr){9-9} \cmidrule(l){10-10}
Seq.       & 75.0$\pm${11} & ~-8.1$\pm${15} & -30.1$\pm${24} & 88.1$\pm${6} & -5.0 & -4.6 & 55.4$\pm${10} & -32.7$\pm${16}    & ~\enspace\underline{-3.9}$\pm${0.2} \\
RWalk     & \underline{77.5}$\pm${ 7} & ~\underline{-6.6}$\pm${13} & -28.3$\pm${26} & 87.4$\pm${7} & -7.1 & -4.6 & 52.5$\pm${ 9} & -35.1$\pm${18}    & ~\enspace\underline{-3.9}$\pm${0.0} \\
EWC       & 44.1$\pm${13} & -18.9$\pm${14} & -22.7$\pm${14} & 88.8$\pm${4} & ~\textbf{0.5} & -4.2 & \underline{85.0}$\pm${ 5} & ~\enspace\textbf{0.0}$\pm${0.1}& -11.1$\pm${15} \\
SegReg    & \hspace{-0.3em}\textbf{79.6}$\pm${ 5} & \hspace{0.2em}\textbf{-5.1}$\pm${ 6} & \underline{-21.6}$\pm${22} & \hspace{-0.2em}\textbf{90.5}$\pm${5} & -5.2 & \underline{-4.1} & 82.1$\pm${ 7} & -10.5$\pm${11}    & \enspace\textbf{-3.6}$\pm${1.3} \\
 +EWC     & 69.3$\pm${ 7} & ~-8.9$\pm${ 3} & \hspace{-0.3em}\textbf{-17.8}$\pm${14} & \underline{89.9}$\pm${4} & ~\underline{0.2} & \textbf{-3.7} & \hspace{-0.2em}\textbf{87.3}$\pm${ 4} & ~\enspace\underline{-0.3}$\pm${0.3}   & \enspace-6.3$\pm${13} \\ \midrule
Reh.       & 82.4$\pm${ 2} & ~-2.4$\pm${ 1} & -23.6$\pm${20} & 91.0$\pm${3} & ~0.2 & -4.9 & 88.6$\pm${ 3} & ~-0.6$\pm${0.2}   & \enspace-4.0$\pm${0.9} \\
\bottomrule
\end{tabular}

    }
\end{table}

Overall continual learning metrics is summarised in Tab.~\ref{tab:cl_metrics} across prostate, cardiac, and hippocampus datasets. SegReg consistently improves forward transfer and final task performance compared to Seq., EWC, and RWalk, while narrowing the gap to rehearsal methods that rely on storing past data. This demonstrates that stabilising latent representations provides an effective and memory-efficient mechanism for mitigating catastrophic forgetting.

\paragraph{Representational Drift Across Tasks.}
\begin{figure}[tb]
    \centering
    \input{figures/pca_w_legend}
    \label{fig:pca}
\end{figure}

To qualitatively assess representational drift, Fig.~\ref{fig:pca} visualises PCA projections of penultimate-layer features across training stages for prostate, cardiac, and hippocampus sequences. Even after the first task, SegReg induces more compact and clearly separated class clusters compared to the baseline, indicating that explicit latent regularisation already improves representation structure in static training. As tasks progress, models without SegReg exhibit noticeable shifts in principal directions and increasing cluster distortion or overlap, reflecting task-induced drift. In contrast, SegReg preserves dominant principal axes and maintains consistent relative positioning of class clusters across stages. This qualitative stability supports our quantitative findings, suggesting that aligning features to a reference distribution constrains representational drift and promotes transferable latent spaces.

\section{Discussion and Conclusion}\label{sec:conc}
In this work, we introduced {SegReg}, a latent-space regularisation framework for medical image segmentation that explicitly structures penultimate-layer embeddings while remaining fully compatible with standard voxel-wise losses. Across prostate, cardiac, and hippocampus datasets within nnU-Net, SegReg consistently improved segmentation accuracy and domain generalisation.

In continual learning, SegReg reduced representational drift, improved forward transfer, and maintained competitive backward transfer without requiring additional memory, parameters, or access to previous task data. These results indicate that constraining latent geometry directly addresses a key limitation of standard segmentation training, where only output predictions are supervised.

Importantly, SegReg is complementary to parameter-based continual learning strategies such as EWC. While EWC constrains weight updates according to parameter importance, it does not explicitly regulate the structure of latent representations. Our results show that combining both approaches further improves stability and forward transfer, suggesting that parameter-space and latent-space regularisation act on distinct but synergistic aspects of the learning dynamics.  Although we focused on domain-incremental continual learning, SegReg is model-agnostic and can be integrated with other regularisation-based strategies. Extending the framework to task-incremental settings and broader continual learning methods remains an interesting direction for future work. 

Overall, our results indicate that structuring the latent space provides benefits beyond standard generalisation: it enables mathematically principled, memory-efficient, and forward-transfer-enhancing continual learning, making SegReg a practical and versatile approach for building more robust medical image segmentation models.


\begin{credits}
\subsubsection{\ackname} This publication is part of the project ROBUST: Trustworthy AI-based Systems for Sustainable Growth with project number KICH3.LTP.20.006, which is (partly) financed by the Dutch Research Council (NWO), Siemens Healthineers, and the Dutch Ministry of Economic Affairs and Climate Policy (EZK) under the program LTP KIC 2020-2023.

\end{credits}

\bibliographystyle{splncs04}
\bibliography{references}

\begin{thebibliography}{10}
\providecommand{\url}[1]{\texttt{#1}}
\providecommand{\urlprefix}{URL }
\providecommand{\doi}[1]{https://doi.org/#1}

\bibitem{balestriero_lejepa_2025}
Balestriero, R., LeCun, Y.: {LeJEPA}: provable and scalable self-supervised learning without the heuristics (Nov 2025). \doi{10.48550/arXiv.2511.08544}

\bibitem{boccardi_training_2015}
Boccardi, M., Bocchetta, M., Frisoni, G.B., et. al.: Training labels for hippocampal segmentation based on the {EADC}-{ADNI} harmonized hippocampal protocol. Alzheimer's \& Dementia: The Journal of the Alzheimer's Association  \textbf{11}(2),  175--183 (Feb 2015). \doi{10.1016/j.jalz.2014.12.002}

\bibitem{campelloMulticentreMultivendorMultidisease2021}
Campello, V.M., Gkontra, P., Lekadir, K., et. al.: Multi-centre, multi-vendor and multi-disease cardiac segmentation: the {M}\&ms challenge. IEEE Transactions on Medical Imaging  \textbf{40}(12),  3543--3554 (Dec 2021). \doi{10.1109/TMI.2021.3090082}

\bibitem{chaitanya_contrastive_2020}
Chaitanya, K., Erdil, E., Konukoglu, E., et. al.: Contrastive learning of global and local features for medical image segmentation with limited annotations. In: {NeurIPS}, 6-12 (Dec 2020), \url{https://dl.acm.org/doi/10.5555/3495724.3496776}

\bibitem{chen_simple_2020}
Chen, T., Kornblith, S., Norouzi, M., Hinton, G.: A simple framework for contrastive learning of visual representations. In: Proceedings of the 37th {International} {Conference} on {Machine} {Learning}. {ICML}'20, vol.~119, pp. 1597--1607. JMLR.org (Jul 2020), \url{https://dl.acm.org/doi/10.5555/3524938.3525087}

\bibitem{consortiumMONAIMedicalOpen2026}
Consortium: {MONAI}: medical open network for {AI} (Jan 2026). \doi{10.5281/zenodo.18408724}

\bibitem{cover_differential_2005}
Cover, T.M., Thomas, J.A.: Differential entropy. In: Elements of information theory, pp. 243--259. John Wiley \& Sons, Ltd (2005). \doi{10.1002/047174882X.ch8}

\bibitem{delangeContinualLearningSurvey2022}
De~Lange, M., Aljundi, R., Tuytelaars, T.: A continual learning survey: defying forgetting in classification tasks. IEEE Transactions on Pattern Analysis and Machine Intelligence  \textbf{44}(7),  3366--3385 (Jul 2022). \doi{10.1109/TPAMI.2021.3057446}

\bibitem{dou_unsupervised_2018}
Dou, Q., Ouyang, C., Heng, P.A., et. al.: Unsupervised cross-modality domain adaptation of convnets for biomedical image segmentations with adversarial loss. In: Proceedings of the 27th {International} {Joint} {Conference} on {Artificial} {Intelligence}. pp. 691--697. {IJCAI}'18, AAAI Press, Stockholm, Sweden (Jul 2018)

\bibitem{gonzalez_lifelong_2023}
González, C., Ranem, A., Mukhopadhyay, A., et. al.: Lifelong {nnU}-{Net}: a framework for standardized medical continual learning. Scientific Reports  \textbf{13}(1), ~9381 (Jun 2023). \doi{10.1038/s41598-023-34484-2}

\bibitem{hadsell_dimensionality_2006}
Hadsell, R., Chopra, S., LeCun, Y.: Dimensionality reduction by learning an invariant mapping. In: 2006 {IEEE} {Computer} {Society} {Conference} on {Computer} {Vision} and {Pattern} {Recognition} ({CVPR}'06). vol.~2, pp. 1735--1742 (Jun 2006). \doi{10.1109/CVPR.2006.100}

\bibitem{he_deep_2016}
He, K., Zhang, X., Ren, S., Sun, J.: Deep residual learning for image recognition. In: 2016 {IEEE} {Conference} on {Computer} {Vision} and {Pattern} {Recognition} ({CVPR}). pp. 770--778 (Jun 2016). \doi{10.1109/CVPR.2016.90}

\bibitem{isensee_nnu-net_2021}
Isensee, F., Jaeger, P.F., Kohl, S.A.A., Petersen, J., Maier-Hein, K.H.: {nnU}-net: a self-configuring method for deep learning-based biomedical image segmentation. Nature Methods  \textbf{18}(2),  203--211 (Feb 2021). \doi{10.1038/s41592-020-01008-z}

\bibitem{kamnitsas_unsupervised_2017}
Kamnitsas, K., Baumgartner, C., Glocker, B., et. al.: Unsupervised domain adaptation in brain lesion segmentation with adversarial networks (May 2017). \doi{10.1007/978-3-319-59050-9_47}

\bibitem{kingma_auto-encoding_2014}
Kingma, D.P., Welling, M.: Auto-encoding variational bayes. In: Bengio, Y., LeCun, Y. (eds.) 2nd {International} {Conference} on {Learning} {Representations}, {ICLR} 2014, {Banff}, {AB}, {Canada}, {April} 14-16, 2014, {Conference} {Track} {Proceedings} (2014), \url{http://arxiv.org/abs/1312.6114}

\bibitem{kirkpatrick_overcoming_2017}
Kirkpatrick, J., Pascanu, R., Rabinowitz, N., Hadsell, R., et. al.: Overcoming catastrophic forgetting in neural networks. Proceedings of the National Academy of Sciences  \textbf{114}(13),  3521--3526 (Mar 2017). \doi{10.1073/pnas.1611835114}

\bibitem{kohl_probabilistic_2018}
Kohl, S.A.A., Romera-Paredes, B., Ronneberger, O., et. al.: A probabilistic {U}-net for segmentation of ambiguous images. In: Proceedings of the 32nd {International} {Conference} on {Neural} {Information} {Processing} {Systems}. pp. 6965--6975. {NIPS}'18, Curran Associates Inc., Red Hook, NY, USA (Dec 2018), \url{https://dl.acm.org/doi/10.5555/3327757.3327800}

\bibitem{kulaga-yoskovitz_multi-contrast_2015}
Kulaga-Yoskovitz, J., Bernhardt, B.C., Bernasconi, N., et. al.: Multi-contrast submillimetric 3-tesla hippocampal subfield segmentation protocol and dataset. Scientific Data  \textbf{2},  150059 (2015). \doi{10.1038/sdata.2015.59}

\bibitem{liuMSnetMultisiteNetwork2020}
Liu, Q., Dou, Q., Yu, L., Heng, P.A.: {MS}-net: multi-site network for improving prostate segmentation with heterogeneous {MRI} data. IEEE transactions on medical imaging  \textbf{39}(9),  2713--2724 (Sep 2020). \doi{10.1109/TMI.2020.2974574}

\bibitem{mccloskey_catastrophic_1989}
McCloskey, M., Cohen, N.J.: Catastrophic interference in connectionist networks: the sequential learning problem  \textbf{24},  109--165 (1989). \doi{10.1016/S0079-7421(08)60536-8}

\bibitem{oktay_anatomically_2018}
Oktay, O., Ferrante, E., Rueckert, D., et. al.: Anatomically constrained neural networks ({ACNNs}): application to cardiac image enhancement and segmentation. IEEE Transactions on Medical Imaging  \textbf{37}(2),  384--395 (Feb 2018). \doi{10.1109/TMI.2017.2743464}

\bibitem{rolnickExperienceReplayContinual2019}
Rolnick, D., Ahuja, A., Wayne, G., et. al.: Experience replay for continual learning. In: {NeurIPS}, pp. 350--360. No.~32, Curran Associates Inc., Red Hook, NY, USA (Dec 2019), \url{https://dl.acm.org/doi/10.5555/3454287.3454319}

\bibitem{ronneberger_u-net_2015}
Ronneberger, O., Fischer, P., Brox, T.: U-net: convolutional networks for biomedical image segmentation. In: Navab, N., Hornegger, J., Wells, W.M., Frangi, A.F. (eds.) Medical {Image} {Computing} and {Computer}-assisted {Intervention} – {MICCAI} 2015. pp. 234--241. Springer International Publishing, Cham (2015). \doi{10.1007/978-3-319-24574-4_28}

\bibitem{shorten_survey_2019}
Shorten, C., Khoshgoftaar, T.M.: A survey on {Image} {Data} {Augmentation} for {Deep} {Learning}. Journal of Big Data  \textbf{6}(1), ~60 (Jul 2019). \doi{10.1186/s40537-019-0197-0}

\bibitem{simpson_large_2019}
Simpson, A.L., Antonelli, M., Cardoso, M.J., et. al.: A large annotated medical image dataset for the development and evaluation of segmentation algorithms. CoRR  (2019). \doi{10.48550/arXiv.1902.09063}

\bibitem{vaish_data-agnostic_2025}
Vaish, P., Meister, F., Wolterink, J.M., et. al.: Data-agnostic augmentations for unknown variations: out-of-distribution generalisation in {MRI} segmentation (Jan 2025), \url{https://openreview.net/forum?id=erHgJGtptZ}

\bibitem{vu_advent_2019}
Vu, T.H., Jain, H., Pérez, P., et. al.: {ADVENT}: adversarial entropy minimization for domain adaptation in semantic segmentation (Apr 2019). \doi{10.48550/arXiv.1811.12833}

\end{thebibliography}

\end{document}